\documentclass[aps,prl,twocolumn,groupedaddress,superscriptaddress,amsfonts,amssymb,amsmath,citeautoscript,a4paper]{revtex4-1}

\usepackage[utf8]{inputenc}
\usepackage[english]{babel}
\usepackage{microtype}
\usepackage{txfonts}
\usepackage{txfontsb}

\usepackage{bm}
\usepackage{xcolor}
\usepackage{graphicx}
\usepackage{float}
\usepackage{xspace}
\usepackage{etoolbox}

\usepackage{physics}
\usepackage{enumerate}
\usepackage[inline]{enumitem}

\usepackage{siunitx}
\usepackage[pdftex]{hyperref}
\hypersetup{colorlinks,
            linkcolor={blue!75!black!80!yellow},
            citecolor={blue!75!black!80!yellow},
            urlcolor={blue!75!black!80!yellow},
            pdfstartview=FitH}

\usepackage[eulergreek]{sansmath}
\makeatletter
\renewcommand\@make@capt@title[2]{%
    \@ifx@empty\float@link{\@firstofone}{\expandafter\href\expandafter{\float@link}}%
    \sisetup{math-sf=\textsf}%
    \sansmath\sffamily\textbf{#1\@caption@fignum@sep}#2
}%

\makeatother

\newcommand{\e}{\mathrm{e}}
\newcommand{\iu}{\mathrm{i}}

\newcommand{\epsb}{\epsilon_\mathrm{b}}

\newcommand{\EF}{E_\mathrm{F}}
\newcommand{\alphafs}{\alpha}
\newcommand{\betaPE}{\beta_{\text{\textsc{PE}}}}

\DeclareMathOperator{\Ei}{Ei}

\newcommand{\appropto}{\mathrel{\vcenter{
			\offinterlineskip\halign{\hfil$##$\cr
				\propto\cr\noalign{\kern.5pt}\sim\cr\noalign{\kern-1pt}}}}}

\newcommand{\ICFOaffil}{ICFO -- Institut de Ci{\`e}ncies Fot{\`o}niques, The Barcelona Institute of Science and Technology, 08860 Castelldefels (Barcelona), Spain}
\newcommand{\ICREAaffil}{ICREA -- Instituci\'o Catalana de Recerca i Estudis Avan\c{c}ats, Passeig Llu\'{\i}s Companys 23, 08010 Barcelona, Spain}

\begin{document}
\title{Multi-plasmon effects and plasmon satellites in photoemission from nanostructures}
%
\author{P.~A.~D.~Gon\c{c}alves}
\affiliation{\ICFOaffil}
\author{F.~Javier~Garc\'{\i}a~de~Abajo}
\email{javier.garciadeabajo@nanophotonics.es}
\affiliation{\ICFOaffil}\affiliation{\ICREAaffil}


\begin{abstract}
Plasmons can be excited during photoemission and produce spectral photoelectron features that yield information on the nanoscale optical response of the probed materials. However, these so-called plasmon satellites have so far been observed only for planar surfaces, while their potential for the characterization of nanostructures remains unexplored. Here, we theoretically demonstrate that core-level photoemission from nanostructures can display spectrally narrow plasmonic features, reaching relatively high probabilities similar to the direct peak. Using a nonperturbative quantum-mechanical framework, we find a dramatic effect of nanostructure morphology and dimensionality as well as universal scaling laws for the plasmon-satellite probabilities. In addition, we introduce a pump--probe scheme in which plasmons are optically excited prior to photoemission, leading to plasmon losses and gains in the photoemission spectra and granting us access to the ultrafast dynamics of the sampled nanostructure. These results emphasize the potential of plasmon satellites to explore multi-plasmon effects and ultrafast electron--plasmon dynamics in metal-based nanoparticles and two-dimensional nanoislands.
\end{abstract}

\maketitle


\section{Introduction}

Photoemission spectroscopies (PESs) hold a central place in materials and atomic science to characterize the electronic structure of solids and molecules~\cite{H03_2}. Prominent examples of PES include x-ray, ultraviolet, and angle-resolved photoemission spectroscopies (XPS, UPS, and ARPES). The fundamental process common to all types of PES is the photoelectric effect---observed by Hertz~\cite{H1887} and explained by Einstein~\cite{E1905} over a century ago---whereby a sufficiently energetic photon is absorbed by an electron that is then ejected from the specimen. Measurement and analysis of the emitted photoelectrons provide chemical, structural, and electronic information about the system. Currently, PESs are consolidated tools in forefront research areas ranging from ultrafast electron dynamics in atoms and molecules~\cite{OSS17,BFO20} to fundamental excitations and electronic processes in condensed matter such as charge transfer~\cite{BKM02,FFH05}, electron transport~\cite{CMU07,NEB12}, dynamic screening~\cite{CTC17,VSS19}, and strongly correlated phenomena~\cite{DHS03}, as well as plasmon-assisted photoemission from flat metal surfaces~\cite{LRW21,NDR22,GK23}, including those supporting quasi-two-dimensional surface states~\cite{MAB19,GNP22}.

Photoemission from a metal surface is not a single-particle process, but rather the result of a complex interplay between the created hole, the outgoing photoelectron, and all other electrons and excitations of the many-body system~\cite{MRC16}. Consequently, the analysis of photoemission spectra provides a window into the interactions and dynamics of electrons and their collective excitations in the sampled material. In particular, plasmons imprint additional features on the photoemission spectra---the so-called {\it plasmon satellites}~\cite{H03_2,MRC16,CVG20}---that accompany a given core-level peak when the photoelectron--hole complex excites these collective modes via Coulomb interaction at the expense of the photoelectron kinetic energy~\cite{H03_2,MRC16,CVG20}.

Following early observations~\cite{KLM1973,PME1975,FBB1977,HSH1977,WWA1977,SHH1978,BWB1979,JL1979,CL1978,LC1979}, plasmon satellites remain an active subject of experimental and theoretical research~\cite{RH05,LPV15,ZKS15,CVP18,ZGK18,SMG18,ZRN20,CR20,RK21}, as their intensity and spectral signatures carry information on correlations and ultrafast dynamic processes that go beyond the static electronic structure. However, the analysis of these spectral features does not, in general, grant us direct quantitative information on the associated correlation effects. Theoretical modeling is typically required to disentangle background contributions as well as to separate core-hole (intrinsic) and photoelectron (extrinsic) effects~\cite{H03_2,MRC16,HMI98,CL72,CL73,SS1974,SS1976,G1977,I1983,H99}. Strikingly, the existing studies of plasmon satellites in PES have so far been restricted to extended planar metal surfaces, while finite-sized metallic nanostructures have not yet been considered despite their potential to yield sharper satellite peaks that capitalize on the dispersionless nature of localized surface plasmons (LSPs) in nanoparticles~\cite{paper300} combined with the superior ability of these excitations to squeeze electromagnetic fields down to deeply subwavelength scales~\cite{MSB14}. 

Here, we theoretically demonstrate the existence of intense and spectrally narrow plasmon satellites in core-level photoemission from nanostructures such as metallic nanoparticles and graphene nanoislands. By introducing a nonperturbative quantum-mechanical formalism that captures the ultrafast dynamics of the photoelectron, the photohole, and the plasmonic boson field, we derive closed-form analytical expressions for the intensities of plasmon satellites associated with the simultaneous excitation of multiple LSPs. Our results reveal a dramatic effect of particle morphology and dimensionality on the satellite probabilities, which we present in the form of material- and size-independent universal scaling laws. We further investigate nanostructures in which the plasmon field has been excited by resonant illumination prior to photoemission, introducing additional satellites at energies above the main core-level peak (i.e., gain satellites) and enabling the study of the ultrafast electron-plasmon dynamics in a pump--probe scheme. Our work underscores the potential of plasmon satellites from nanostructures for exploring multi-plasmon effects, ultrafast plasmon dynamics, and electron--plasmon interactions in metallic nanoparticles and two-dimensional nanoislands.

\begin{figure*}[t]
\centering
\includegraphics[width=0.65\textwidth]{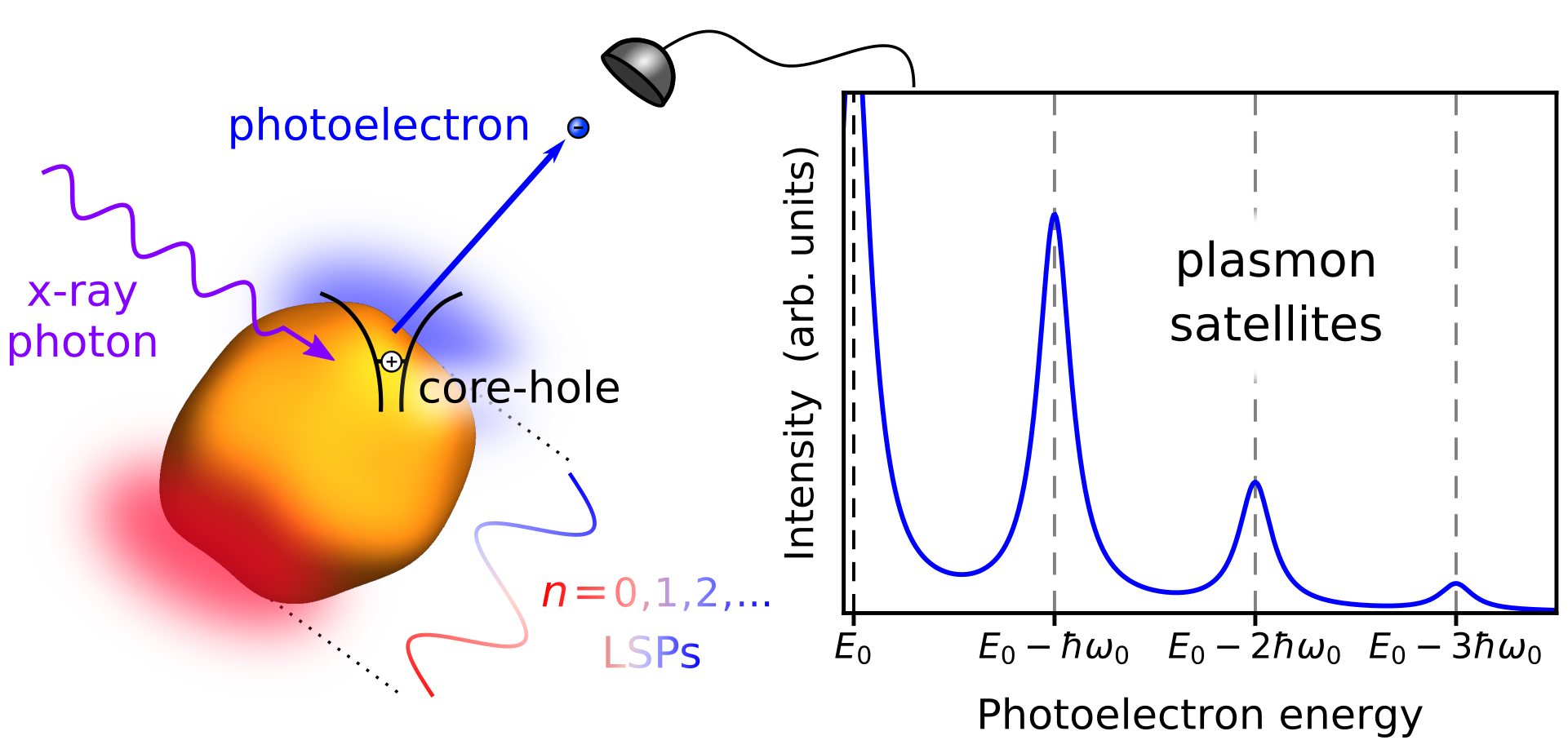}
\caption{\textbf{Plasmon satellites in the photoemission spectrum from a metallic nanoparticle.} A sufficiently energetic photon ejects an electron from a deep electronic level, leaving a localized core hole behind. Dynamic screening of the hole and photoelectron by conduction electrons in the metal gives rise to the excitation of localized surface plasmons (LSPs), resulting in a photoemission spectrum that exhibits a series of peaks separated by multiples of the LSP energy $\hbar\omega_0$ from the main core-level peak (at an electron kinetic energy $E_0$).
}\label{fig:Fig1}
\end{figure*}

\section{Results and discussion}

\subsection{Theoretical framework} 
We consider a particle hosting a dominant plasmon mode, described as a quantum harmonic oscillator~\cite{PN2000,GL91,AMG10,H20} through the free Hamiltonian
\begin{subequations}
\label{eq:H_def}
\begin{align}
\label{eq:Hamiltonian_0}
\hat{\mathcal{H}}_0 &= \hbar \omega_0 \, \hat{a}^\dagger \hat{a},
\end{align}
where $\hat{a}^\dagger$ and $\hat{a}$ are the corresponding bosonic creation and annihilation operators. We treat the photoelectron as a classical particle that follows a straight-line trajectory from the position of the initial localized core level to the detector. The electron--plasmon interaction is then encapsulated in a time-dependent coupling coefficient $g(t)$, entering the Hamiltonian $\hat{\mathcal{H}} = \hat{\mathcal{H}}_0 + \hat{\mathcal{H}}_1$ through a term~\cite{LS1972,LKB1970,FE1985,paper228}
\begin{align}
\label{eq:Hamiltonian_1}
\hat{\mathcal{H}}_1 &= g^*(t) \, \hat{a}^\dagger + g(t) \, \hat{a}.
\end{align}
\end{subequations}
More precisely~\cite{paper228},
\begin{align}
g(t) = \int \mathrm{d}\mathbf{r}\, \phi_{\mathrm{p}}(\mathbf{r}) \, \rho_{\text{ext}}(\mathbf{r},t),
\label{eq:def_g}
\end{align}
where $\phi_{\mathrm{p}}(\mathbf{r})$ denotes the electric potential corresponding to a single plasmon, and $\rho_{\mathrm{ext}}(\mathbf{r},t) = e\,\big[\delta(\mathbf{r}_0-\mathbf{r})-\delta(\mathbf{r}_0+\mathbf{v}t-\mathbf{r}) \big]\,\Theta(t)$ is the charge-density due to both the core hole (first term, assumed to be static and localized at $\mathbf{r}_0$) and the outgoing photoelectron (second term, moving away from the sampled structure with velocity $\mathbf{v}$). The Heaviside step function $\Theta(t)$ limits the interaction to times $t>0$ (i.e., after the photoemission event). For this charge-density, the coupling coefficient in Eq.~\eqref{eq:def_g} reduces to $g(t) = \big[ g_0 + g_1(t) \big]\,\Theta(t)$, with $g_0 \equiv e \phi_{\mathrm{p}}(\mathbf{r}_0)$ and $g_1(t) \equiv - e \phi_{\mathrm{p}}(\mathbf{r}_0 + \mathbf{v}t)$, embodying the contributions from the photohole and the photoelectron, respectively.

The Hamiltonian in Eqs.~\eqref{eq:H_def} is formally equivalent to that of a classically driven quantum harmonic oscillator, whose solution is well-known~\cite{CN1965}. In particular, the time-dependence of the wave function is given by $\ket{\psi(t)}=\hat{S}(t)\ket{\psi(0)}$ in terms of the time-evolution operator $\hat{S}(t) = \e^{\iu \chi(t)}\, \e^{\beta^*(t) \hat{a}^\dagger - \beta(t) \hat{a}}$. The latter is characterized by a single coupling parameter $\beta(t) = \iu \hbar^{-1} \int_0^{t} \mathrm{d}t'\, g(t')\, \e^{-\iu \omega_0 t'}$, along with a global phase $\chi(t)$ that is irrelevant for the present study [see ESI$^\dag$ for further details].

To calculate the transition probabilities associated with the plasmon satellites, it is necessary to determine the eigenstates of the Hamiltonian $\hat{\mathcal{H}}_{\infty} = \hat{\mathcal{H}}_0 + g^*_0 \hat{a}^\dagger + g_0 \hat{a}$ long after the photoemission event (i.e., at $t \to \infty$). This Hamiltonian, which contains the interaction between the plasmons and the core hole left behind, can be diagonalized by performing the unitary transformation~\cite{M00,L1970} $\ket{\tilde{n}} = \e^{-\hat{s}} \ket{n}$, where $\hat{s} = \Delta_0^* \hat{a}^\dagger - \Delta_0 \hat{a}$ with
\begin{align}
\label{eq:Delta0}
\Delta_0 \equiv g_0/\hbar\omega_0 = e \phi_{\mathrm{p}}(\mathbf{r}_0)/\hbar\omega_0.
\end{align}
We find (see SM)  
\begin{align}
\label{eq:H_infty_n-tilde}
\hat{\mathcal{H}}_{\infty} &\ket{\tilde{n}} = \hbar\omega_0 \big( n - |\Delta_0|^2 \big) \ket{\tilde{n}},
\end{align}
and hence $\ket{\tilde{n}}$ are the eigenstates of $\hat{\mathcal{H}}_{\infty}$ with eigenenergies $\hbar\omega_0 \big( n - |\Delta_0|^2 \big)$ [i.e., those of the (bare) quantum harmonic oscillator, but globally shifted by a constant energy $-|g_0|^2/\hbar\omega_0$]. 

In what follows, we apply this formalism to describe plasmon satellites in core-level photoemission from finite-sized metal nanoparticles (Fig.~\ref{fig:Fig1}) and nanographenes. For simplicity, we assume that the photoelectrons originate near the particle surface, which is reasonable for kinetic energies $E_0=$\SIrange{10}{1500}{\eV}, whose associated electron escape depths are $\lesssim \SI{1}{\nm}$~\cite{H03_2,SD1979}. Incidentally, the electron kinetic energy associated with a main core-level, $E_0 = h\nu_{X} - |E_{\text{\textsc{B}}}| - |\Delta_0|^2 \hbar\omega_0$, is determined by the x-ray photon energy $h\nu_{X}$, the bare binding energy $E_{\text{\textsc{B}}}$, and a contribution due to core--plasmon interaction $-|\Delta_0|^2\hbar\omega_0$.

We first consider a conventional scenario in which the plasmon state is initially prepared in the ground state $\ket{\psi(0)}=\ket{0}$ (i.e., no plasmons are present before the interaction, which is an excellent approximation under the common condition $\hbar\omega_0 \gg k_{\text{\textsc{B}}}T$, where $T$ is the temperature). The photoemission spectrum for kinetic energies $E$ around the main core-level peak $E_0$ is then given by ${ \Gamma(E) = \sum_{\ell=-\infty}^{0} P_\ell\, \delta(E-E_0 - \ell \hbar\omega_0) }$, where (see ESI$^\dag$ for a self-contained derivation)
\begin{align}
P_\ell = \frac{|\betaPE|^{2|\ell|}}{|\ell|!} \e^{-|\betaPE|^{2}},
\label{eq:Pl_n0_0}
\end{align}
is the probability associated with a plasmon satellite peak corresponding to the excitation of $|\ell|$ LSPs (i.e., a photoelectron energy loss $E_0-E=-\ell\hbar\omega_0$ relative to the direct peak), characterized by a single coupling parameter
\begin{align}
 \betaPE = \Delta_0 + \frac{e}{\iu \hbar} \int_{0}^{\infty} \phi_{\mathrm{p}}(\mathbf{r}_0 + \mathbf{v}t)\, \e^{-\iu \omega_0 t} \mathrm{d}t.
 \label{eq:betaPE}
\end{align}
The plasmon satellite intensities follow a Poissonian distribution with an average number of excited plasmons given by $|\betaPE|^2$. The first term in Eq.~(\ref{eq:betaPE}) entails the contribution from the core hole, whereas the second one accounts for the interaction with the photoelectron and depends on the orientation and magnitude of the velocity $\mathbf{v}$ (e.g., it vanishes in the large-velocity limit). Incidentally, the coupling parameter in Eq.~(\ref{eq:betaPE}) can be rewritten in terms of the plasmon electric field $\mathbf{E}_\mathrm{p}(\mathbf{r})=-\nabla\phi_{\mathrm{p}}(\mathbf{r})$ as (see ESI$^\dag$)
\begin{align}
\betaPE = \frac{e}{\hbar \omega_0} \int_{z_0}^{\infty} E_{\mathrm{p},z}(z)\, \e^{-\iu \omega_0 (z-z_0)/v} \, \mathrm{d}z,
\label{eq:betaPE_Ez}
\end{align}
where we have taken the electron velocity along $z$ without loss of generality. The coupling strength in Eq.~(\ref{eq:betaPE_Ez}) amounts to the Fourier transform of the plasmon electric-field component along the photoelectron trajectory, with the probed spatial frequency $\omega_0/v$ dictated by the plasmon frequency $\omega_0$ and the electron escape velocity $v$. This result is analogous to the coupling coefficient between free electrons and LSPs in electron energy-loss spectroscopy (EELS)~\cite{paper149}, cathodoluminescence (CL)~\cite{paper338}, and photon-induced near-field electron microscopy (PINEM)~\cite{BFZ09,paper151,FES15,paper311}, where the lower limit of integration is instead extended to $z_0=-\infty$.

\begin{figure*}[t]
\centering
\includegraphics[width=1.0\textwidth]{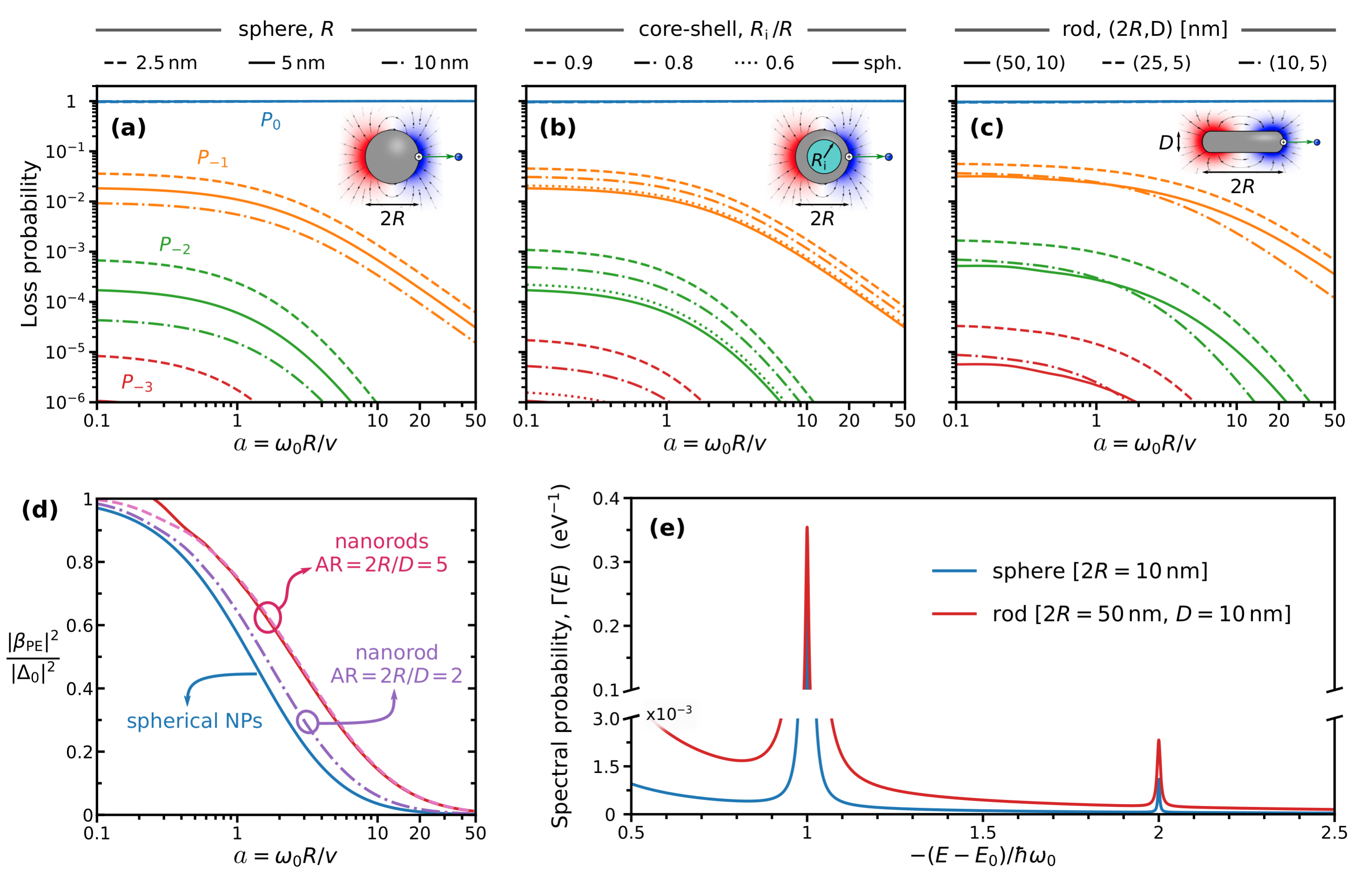}
\caption{\textbf{Plasmon satellites in core-level photoemission from metal nanoparticles.} \textbf{(a)--(c)}~Probabilities of observing satellites associated with multiple LSP excitations in Ag nanoparticles of different morphologies: \textbf{(a)} nanospheres, \textbf{(b)} spherical nanoshells ($R=\SI{5}{\nm}$), and \textbf{(c)} nanorods with hemispherical ends. The direction of emission is parallel to the mode polarization, as indicated in the insets. Probabilities are normalized to that of the direct core-level peak in the absence of plasmons. The probability $P_\ell$ of exciting $|\ell|$ plasmons is calculated for different geometrical parameters (sphere radius in nanospheres, inner-to-outer radius ratio in nanoshells, length and diameter in rods), as indicated in the upper legends. \textbf{(d)} Material- and size-independent universal curves characterizing the squared modulus of the photoemission parameter for subwavelength metallic nanoparticles. \textbf{(e)} Photoelectron energy-loss spectrum (measured with respect to the zero-plasmon line at energy $E_0 = \SI{1}{\keV}$) from a Ag nanosphere (blue, radius $R = \SI{5}{\nm}$) and from a Ag nanorod (red, length $2R = \SI{50}{\nm}$ and diameter $D=\SI{10}{\nm}$). We model the permittivity of silver as $\epsilon_{\mathrm{m}}(\omega)=\epsb-\omega_{\mathrm{p}}^2/\omega(\omega+\iu\gamma)$ with parameters $\epsb=4$, $\hbar\omega_{\mathrm{p}}=\SI{9.17}{\eV}$, and $\hbar\gamma=\SI{21}{\meV}$ extracted from optical measurements~\cite{JC1972}. A FWHM Lorenztian broadening equal to $\gamma$ has been introduced in the spectra plotted in (e). The LSP energies are $\SI{3.74}{\eV}$ in the sphere (independent of size in the quasistatic regime); 2.53, 3.20, and 3.55~eV in the shells (in order of decreasing $R_{\text{i}}/R$); 1.91, 1.94, and 3.04~eV in the rods (in order of decreasing length). 
}\label{fig:Fig2}
\end{figure*}

\subsection{Plasmon satellites in core-level photoemission from metal nanoparticles} 
For concreteness, we consider metal nanoparticles whose optical response for polarization along $z$ is dominated by a single LSP mode of dipolar character. We study spherical particles and nanorods of small size compared to the light wavelength at the plasmon frequency. The particle permittivity is approximated by the Drude-like expression $\epsilon_{\mathrm{m}}(\omega)=\epsb-\omega_{\mathrm{p}}^2/\omega(\omega+\iu\gamma)$, with parameters $\epsb=4$, $\hbar\omega_{\mathrm{p}}=\SI{9.17}{\eV}$, and $\hbar\gamma=\SI{21}{\meV}$ describing Ag~\cite{paper300}, as deduced from optical measurements~\cite{JC1972}. For spherical particles, the single-plasmon electric potential can be approximated as $\phi_\mathrm{p}(\mathbf{r}) = \mathbf{p} \cdot \mathbf{r}/r^3$ in terms of the electric dipole moment $\mathbf{p}$ normalized to a single LSP quantum. Further assuming that $\mathbf{p}$, $\mathbf{v}$, and $\mathbf{r}_0$ are all aligned along $z$, and $\mathbf{r}_0 = R \mathbf{\hat{z}}$ is placed at the particle surface, the parameter in Eq.~(\ref{eq:Delta0}) reduces to (see ESI$^\dag$)
\begin{align}
\Delta_0 = \sqrt{\dfrac{3\alphafs/2}{\epsb+2}\,\dfrac{c}{R\,\omega_0 }},
\nonumber
\end{align}
where $\alphafs\simeq 1/137$ is the fine-structure constant. Moreover, inserting the aforementioned dipolar potential into Eq.~(\ref{eq:betaPE}), we find the closed-form analytical expression ${ \betaPE = \Delta_0 \left\{ 1 - \iu a + a^2\, \e^{\iu a} \left[ \iu \pi + \Ei(-\iu a) \right] \right\} }$ for the coupling parameter in terms of the dimensionless quantity $a = \omega_0 R/v$ and the exponential integral function $\Ei$~\cite{AS1972}. We also present results below for nanorods, in which $\betaPE$ is obtained upon numerical integration of Eq.~(\ref{eq:betaPE_Ez}) with the plasmon field ${\bf E}_\mathrm{p}$ approximated by the induced near field displayed under resonant illumination, which is in turn calculated by using the boundary-element method (BEM) \cite{paper040}. The latter is automatically accounting for retardation effects in the mode field.

Figures~\ref{fig:Fig2}a--c show the plasmon satellite probabilities $P_\ell$ calculated from Eq.~\eqref{eq:Pl_n0_0} for core-level photoemission from metallic nanospheres (Fig.~\ref{fig:Fig2}a), spherical nanoshells (Fig.~\ref{fig:Fig2}b), and nanorods (Fig.~\ref{fig:Fig2}c). The probabilities drop by more than one order of magnitude for each subsequent $\ell$ because $|\betaPE|^2 \ll 1$ (i.e., we are in the weak-coupling regime). Also, they are monotonically decreasing functions of the dimensionless parameter $a=\omega_0 R/v$, such that $P_\ell$ is maximum at $a \to 0$ [the so-called sudden limit~\cite{H03_2,MRC16}, wherein the photoelectron is ejected with infinite velocity, and thus, the interaction is solely due to dynamical screening of the core hole (intrinsic contribution)]. As $a$ increases, the contribution due to the outgoing photoelectron [second term in Eq.~\eqref{eq:betaPE}, extrinsic contribution] interferes with that stemming from the photohole in a destructive fashion, thereby reducing $P_\ell$, and eventually producing an exact cancellation in the $a \to \infty$ limit.

The results presented in Figs.~\ref{fig:Fig2}a--c also shed light on the dependence of $P_\ell$ on particle geometry. For spherical particles, smaller sizes lead to more intense plasmon satellites (Fig.~\ref{fig:Fig2}a) as a result of the larger field confinement (i.e., $|\Delta_0|^2 \propto R^{-1}$). In addition, when moving from the nanosphere to a core-shell geometry, the plasmon energy can be tuned through the shell thickness, while spatial confinement near the metal region together with the redshifted resonant frequency results in higher probabilities $P_\ell$ (Fig.~\ref{fig:Fig2}b). In the quasistatic limit, the shell plasmon energy only depends on the ratio of inner-to-outer radii (see ESI$^\dag$) and thinner shells yield higher plasmon satellite probabilities (Fig.~\ref{fig:Fig2}b). Thus, the core-shell geometry can be exploited in this way to counteract the size-dependent decrease of satellite probabilities observed in larger nanoparticles. An analogous effect is also observed in nanorods (see Fig.~\ref{fig:Fig2}c), as plasmon satellite probabilities associated with larger rods can be higher or comparable to those of smaller rods provided that the aspect ratio of the former is higher [cf. (50,10) and (10,5)]. This can be intuitively understood by noting that elongated rods exhibit LSPs shifted towards the red while still yielding similar field confinement, which, together, contribute to larger $\Delta_0=e\phi_{\mathrm{p}}(R)/(\hbar\omega_0)$ and higher satellite probabilities. In practice, for rod lengths $\gtrsim \SI{100}{\nm}$ retardation effects eventually lead to a weakening of the field confinement~\cite{paper300}.

Interestingly, for deeply subwavelength nanoparticles ($R \ll \lambda_0$), the ratio $|\betaPE|^2/|\Delta_0|^2$ is rendered both material- and size-independent, leading to the universal curves shown in Fig.~\ref{fig:Fig2}d, which only depend on the morphology and aspect ratio (AR). As representative examples of our findings, we plot in Fig.~\ref{fig:Fig2}e photoemission spectra from  two silver nanoparticles, namely, a nanosphere (blue line) and a nanorod (red line), in which two discernible plasmon satellites can be identified, associated with the excitation of one and two LSP quanta, though the latter is substantially weaker. Incidentally, a size dependence is observed due to retardation when comparing the two nanorods of AR equal to 5 in Fig.~\ref{fig:Fig2}d (see coordinated legend in Fig.~\ref{fig:Fig2}c).

\begin{figure*}[t]
\centering
\includegraphics[width=1.0\textwidth]{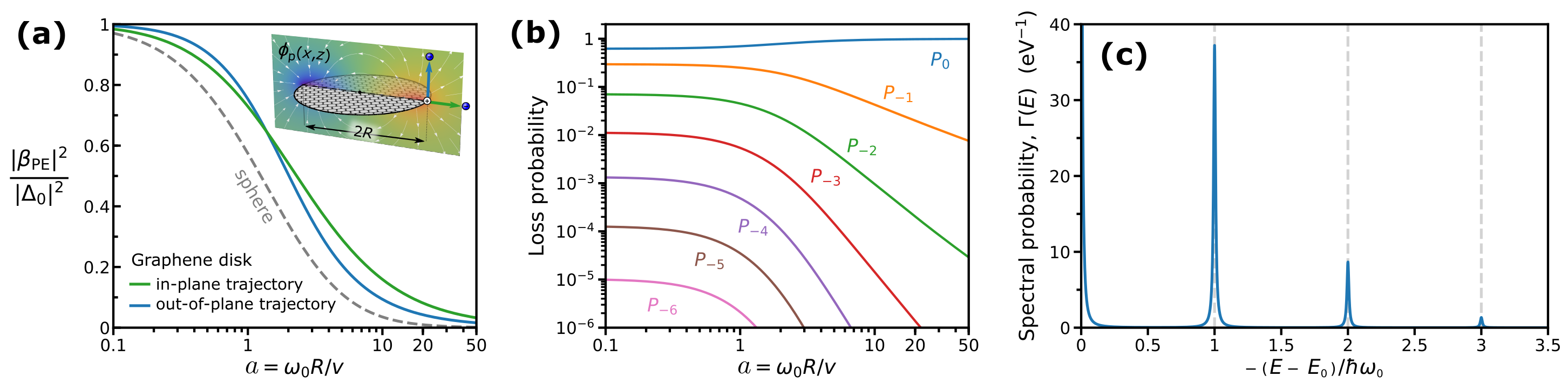}
\caption{\textbf{Plasmon satellites in photoemission from graphene nanodisks.} \textbf{(a)}~Size-independent universal curves of the absolute square of the normalized photoemission parameter in graphene disks, corresponding to the color-matched trajectories indicated in the inset. The latter shows a graphene nanodisk along with its dipole-plasmon potential and field lines. \textbf{(b)}~Plasmon satellite probabilities $P_\ell$ for excitation of $|\ell|$ LSPs in a graphene nanodisk with radius $R = \SI{10}{\nm}$ and Fermi energy $E_{\mathrm{F}} = \SI{0.25}{\eV}$ (plasmon energy $\hbar\omega_0 = \SI{0.28}{\eV}$). \textbf{(c)}~Photoemission energy-loss spectrum (relative to the main core-level line with a kinetic energy $E_0 = \SI{1}{\keV}$) containing several plasmon satellites. Graphene is described by the Drude surface conductivity $\sigma(\omega) = (\iu e^2E_{\mathrm{F}}/\pi \hbar^2)/(\omega + \iu\gamma)$ with $\hbar\gamma = \SI{5}{\meV}$. A FWHM Lorentzian broadening equal to $\gamma$ has been introduced in (c). 
}\label{fig:Fig3}
\end{figure*}

\subsection{Plasmon satellites from two-dimensional nanostructures} 
Two-dimensional polaritonics has recently become a prominent research field~\cite{paper283} where, in particular, highly-doped graphene nanoislands are known to host highly-confined gate-tunable mid-infrared plasmons~\cite{GP16,paper235}. Stimulated by this, we study satellites in photoemission from self-standing graphene nanodisks (Fig.~\ref{fig:Fig3}a, inset), whose small size relative to the optical wavelength justifies the use of the electrostatic limit. Specifically, for a radius $R$ and Fermi energy $\EF$, the dipolar plasmon energy is $\hbar\omega_0 = \gamma_0 \sqrt{\alphafs \hbar c \EF/2R}$, where $\gamma_0 = 2.088$ is a geometry-dependent eigenvalue~\cite{paper228}. Also, in the long-wavelength limit the single-plasmon potential $\phi_{\mathrm{p}}(\mathbf{r})$ is expressed in terms of a scale-invariant potential distribution (Fig.~\ref{fig:Fig3}a, inset) numerically obtained from BEM~\cite{paper228}. Analogously to the silver nanoparticles discussed above, we plot in Fig.~\ref{fig:Fig3}a universal curves of $|\betaPE|^2/|\Delta_0|^2$ for electrons emitted from the disk edge along either the mode dipole-moment direction or an out-of-plane trajectory (see inset), both showing similar trends as found for the sphere (dashed curve reproduced from Fig.~\ref{fig:Fig2}d). However, due to the stronger confinement in graphene nanodisks relative to that in metallic spheres, the satellite probabilities are about two orders of magnitude larger (cf. Fig.~\ref{fig:Fig2}a--c and Fig.~\ref{fig:Fig3}b). A comb of multi-plasmon excitations is thus predicted to be observable in the photoemission spectra from graphene nanodisks (Fig.~\ref{fig:Fig3}c). We remark that these results are scale-invariant, so that narrow disks (small thickness-to-radius ratio) should exhibit similar combs with probabilities given by Eq.~(\ref{eq:Pl_n0_0}), with the parameter $|\betaPE|^2/|\Delta_0|^2$ depending on $a=\omega_0 R/v$ as shown in Fig.~\ref{fig:Fig3}a, and a plasmon--core-hole interaction scaling as $|\Delta_0|^2 \propto (E_{\mathrm{F}} R)^{-1/2}$.

\begin{figure*}[t]
\centering
\includegraphics[width=1.0\textwidth]{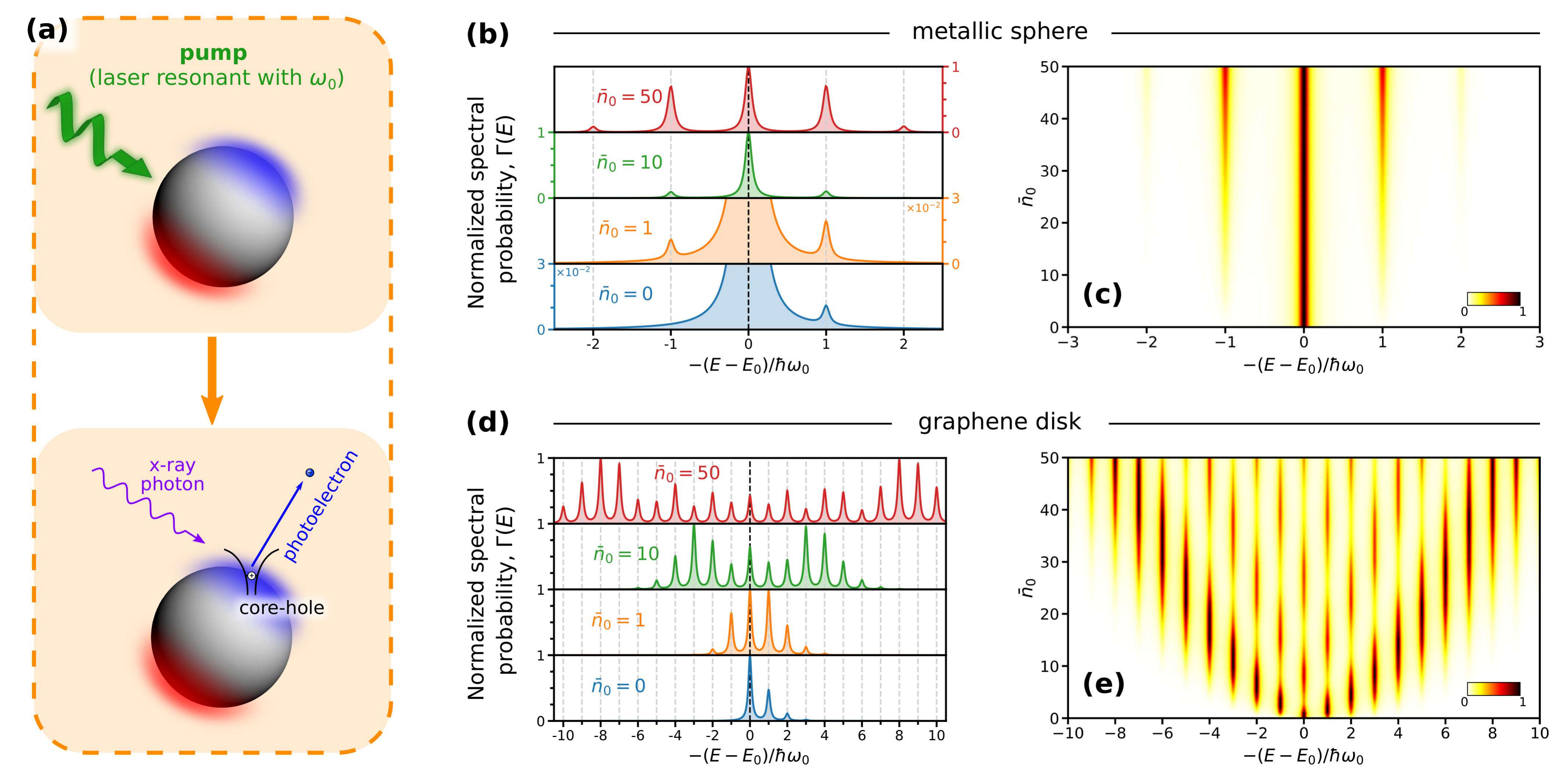} 
\caption{\textbf{Gain and loss plasmon satellites in photoemission from an optically pumped nanoparticle.} \textbf{(a)}~A plasmonic nanostructure is resonantly pumped, thereby exciting the targeted plasmon mode to a coherent state (top). Photoemission from a core level is subsequently produced upon irradiation with x-rays (bottom). 
\textbf{(b)--(c)}~Photoemission energy-loss spectra from a silver nanosphere (radius $R=\SI{5}{\nm}$, LSP energy $\SI{3.74}{\eV}$) hosting a dipolar resonance in a coherent state with an average number of plasmons $\bar{n}_0$. The direction of emission is taken along the plasmon dipole (see inset in Fig.~\ref{fig:Fig2}a).
\textbf{(d)--(e)}~Same as (b)--(c), but now for a graphene nanodisk (radius $R = \SI{10}{\nm}$, Fermi energy $E_{\mathrm{F}} = \SI{0.25}{\eV}$, plasmon energy $\hbar\omega_0 = \SI{0.28}{\eV}$). Spectra for each $\bar{n}_0$ are normalized to the maximum probability, and we have introduced a FWHM Lorentzian broadening of $0.1\omega_0$ and $0.2\omega_0$ in the Ag and graphene spectra, respectively).
}\label{fig:Fig4}
\end{figure*}

\subsection{Loss and gain satellites from optically pumped structures} 
Insight into plasmon dynamics can be obtained in a conceptually unexplored scenario consisting in measuring satellites from particles that are optically pumped by a laser on resonance with the plasmon, such that the latter in prepared in a coherent state prior to photoemission. The initial plasmonic state can be written as $\ket{\psi(0)} = \sum_{n_0} c_{n_0} \ket{n_0}$ with coefficients satisfying the Poisson distribution $|c_{n_0}|^2 = \e^{-\bar{n}_0}\, {\bar{n}_0}^{n_0}/n_0!$ for an average population $\bar{n}_0$. The photoemission spectrum $\Gamma(E) = \sum_{\ell=-\infty}^{\infty} P_\ell \, \delta(E-E_0 - \ell\hbar\omega_0)$ is then comprising loss ($\ell<0$) and gain ($\ell>0$) peaks at kinetic energies $E_0+\ell\hbar\omega_0$, with probabilities given by (see ESI$^\dag$)
\begin{subequations}
\label{eq:Pl_illuminated}
\begin{align}
 P_\ell = \sum_{n=\mathrm{max}\{0,-\ell\}}^{\infty} \left|f_\ell^n \right|^2 ,
\end{align}
where
\begin{align}
 \left|f_\ell^n \right|^2 = \left| c_{n+\ell} \ S_{n,n+\ell}(\betaPE) \right|^2
\end{align}
\end{subequations}
is expressed in terms of the analytical matrix elements $S_{n,n'}(\betaPE) \equiv \bra{n}\hat{S}\ket{n'}$ (see explicit expressions in ESI$^\dag$) that depend only on $\betaPE$ [Eq.~(\ref{eq:betaPE})]. This result is derived by taking into account that the number of excitations in the photoelectron--sample system ($n_0 = n + \ell$) is conserved for each initial $n_0$ component~\cite{paper339,paper360}. Incidentally, while here we focus on initial coherent states, the result in Eq.~\eqref{eq:Pl_illuminated} is general for any initial plasmon distribution $|c_{n_0}|^2$.

Figure~\ref{fig:Fig4} shows the evolution of the photoemission spectra from a plasmonic nanosphere (Figs.~\ref{fig:Fig4}b--c) and from a graphene nanodisk (Figs.~\ref{fig:Fig4}d--e) as a function of the initial average number of plasmons $\bar{n}_0$, which is in turn proportional to the pump intensity. Starting from an asymmetric spectrum without gain peaks at $\bar{n}_0=0$, where the {\it elastic} ($\ell=0$) peak is only marginally depopulated due to single- and double-plasmon processes, we then observe the emergence of multiple quanta processes along with gain sidebands (resulting from plasmon absorption by the photoelectron) as $\bar{n}_0$ increases, while the spectrum progressively becomes more symmetric (see Figs.~\ref{fig:Fig4}b and ~\ref{fig:Fig4}d). Indeed, for $\bar{n}_0 |\betaPE|^2 \gtrsim 1$ the elastic, {\it zero-loss} peak starts to be significantly depleted (i.e., its spectral weight is transferred to the lowest-order plasmon satellites). For a given value of $\bar{n}_0$, this effect is substantially more pronounced in the graphene disk because of its large value of $|\betaPE|$ compared with the nanosphere. Finally, for $\bar{n}_0 |\betaPE|^2 \gg 1$ (Figs.~\ref{fig:Fig4}d--e), the spectral weight is dominantly placed in high-order LSP satellites, with the photoelectron spectra exhibiting significant modulation arising from multipath interference of LSP loss and gain processes contributing to the same final channel $\ell$, which is responsible for a quantum-billiard structure analogous to multilevel Rabi oscillations, as also observed in PINEM~\cite{paper151,FES15}.

\section{Discussion}

In summary, we have theoretically demonstrated that localized plasmon resonances can produce spectrally narrow and relatively intense satellites in core-level photoemission spectra from nanostructures. We base these results on a nonperturbative quantum-mechanical theoretical formalism that extends previous methods used in electron microscope spectroscopies~\cite{paper228,LKB1970,paper339}.

The plasmon-corrected photoemission spectra exhibit a Poissonian distribution of plasmon satellites in the absence of pumping, described by a single parameter $\betaPE$ that captures the interaction of the core-hole--photoelectron complex with the single-plasmon field. The particle geometry has a strong effect on the resulting intensities, as captured by universal curves for the $|\betaPE|^2/|\Delta_0|^2$ ratio as a function of the dimensionless parameter $a=\omega_0 R/v$. This constitutes a practical and important result because, for a given particle morphology, $|\betaPE|^2$ can be determined for any particle size and composition, provided that $\Delta_0 = e\phi_{\mathrm{p}}(\mathbf{r}_0)/(\hbar\omega_0)$ is known, thereby facilitating the computation and modeling of plasmon satellites in photoemission from plasmonic nanostructures. In addition, we find that two-dimensional nanoislands lead to particularly intense multi-plasmon satellite peaks due to their ability to confine the mode field distribution down to extremely subwavelength volumes. Furthermore, we predict that, when the plasmon field is populated prior to photoemission by using resonant illumination, energy-gain satellites show up in the spectra, which undergo a complex evolution as a function of pumping intensity, suggesting new possibilities for studying plasmon dynamics in a pump--probe fashion in which the probe is provided by the x-ray exciting the core level and the ensuing photoelectron. For example, controlling the delay between the optical pump and the x-ray triggering photoemission could grant us access to time-resolved plasmon dynamics. More generally, targeting different core levels and/or LSP resonances could be leveraged to probe the underpinning ultrafast plasmon dynamics at different time scales, since the electron--plasmon \emph{effective} interaction time, $\sim L_\mathrm{p}/v$, depends both on the photoelectron escape velocity $v$ and on the extension of the LSP near-field $L_\mathrm{p}$.

We note that the present results can directly be applied to satellites associated with other types of particle excitations, such as phonon-polaritons (e.g., in hexagonal boron nitride nanocrystals) and exciton-polaritons (e.g., in transition metal dichalcogenides). A vast range of possibilities is opened by the fact that the core level under consideration can belong to one of the surface atoms in the structure, an extrinsic (e.g., absorbed) impurity, or an atom in a nearby material (e.g., a substrate). Chemical shifts in the energy of core levels in surface and sub-surface atoms also suggest an approach to probe plasmon fields with a resolution limited by the atomic-bond distances. We note that the selected core level must be narrower than the probed-mode energy, such that satellites are spectrally separated from the direct photoemission peaks. In brief, our work introduces a disruptive approach to explore polariton dynamics in nanostructures spanning a wide range of spectral regions and material platforms.

\hfill

\section*{Acknowledgements}
We thank Enrique Ortega, Valerio Di Giulio, and Andrea Kone\v{c}n\'{a} for stimulating discussions. This work has been supported in part by the European Research Council (Advanced Grant 789104-eNANO), the European Commission (Horizon 2020 Grants 101017720 FET-Proactive EBEAM and 964591-SMART-electron), the Spanish MICINN (PID2020-112625GB-I00 and Severo Ochoa CEX2019-000910-S), the Catalan CERCA Program, and Fundaci\'{o}s Cellex and Mir-Puig.


\providecommand*{\mcitethebibliography}{\thebibliography}
\csname @ifundefined\endcsname{endmcitethebibliography}
{\let\endmcitethebibliography\endthebibliography}{}

\end{document}